\documentclass[aip,reprint,author-numerical]{revtex4-2}

\usepackage{dcolumn}
\usepackage[utf8]{inputenc}
\usepackage[T1]{fontenc}
\usepackage{mathptmx}
\usepackage{caption}

\usepackage{CJK}
\usepackage{graphicx}
\usepackage[justification=justified,singlelinecheck=false]{caption}
\usepackage{bm}
\usepackage{amssymb}
\usepackage{amsmath}
\usepackage[colorlinks,linkcolor=blue,anchorcolor=blue,citecolor=blue]{hyperref}
\usepackage{xcolor}

\makeatletter
\def\@email#1#2{%
 \endgroup
 \patchcmd{\titleblock@produce}
  {\frontmatter@RRAPformat}
  {\frontmatter@RRAPformat{\produce@RRAP{*#1\href{mailto:#2}{#2}}}\frontmatter@RRAPformat}
  {}{}
}
\makeatother

\begin{document}

\title{Lifting spin degeneracy in rhombohedral trilayer graphene for high magnetoresistance applications}
\author{Lishu Zhang\textsuperscript{*}}

\email{lishu.zhang@sdu.edu.cn}
\affiliation{Key Laboratory for Liquid-Solid Structural Evolution and Processing of Materials, Ministry of Education, Shandong University, Jinan 250061, China}

\author{Jun Zhou}
\affiliation{Future Energy Acceleration \& Translation (FEAT), 
Strategic Research \& Translational Thrust (SRTT), 
A*STAR Research Entities, 
1 Fusionopolis Way \#20-10 Connexis North Tower,
Singapore 138632, Republic of Singapore}
\affiliation{Institute of Materials Research and Engineering (IMRE),
Agency for Science, Technology and Research (A*STAR),
2 Fusionopolis Way, Innovis \#08-03,
Singapore 138634, Republic of Singapore.}

\author{Jie Yang}
\affiliation{Key Laboratory of Material Physics, School of Physics, Ministry of Education, Zhengzhou University, Zhengzhou 450001, China}

\author{Sumit Ghosh}
\affiliation{The Institute of Physics of the Czech Academy of Sciences, 162 00 Prague, Czech Republic}

\author{Yi-Ming Zhao}
\affiliation{Department of Mechanical Engineering, National University of Singapore, 9 Engineering Drive 1, Singapore, 117575, Republic of Singapore}

\author{Yuan Ping Feng}
\affiliation{Department of Physics, National University of Singapore, Singapore 117542, Republic of Singapore}

\author{Lei Shen}
\affiliation{Department of Mechanical Engineering, National University of Singapore, 9 Engineering Drive 1, Singapore, 117575, Republic of Singapore}


\begin{abstract}

Many exotic properties in rhombohedral (or ABC-stacked) multilayer graphene have recently been reported experimentally. In this Letter, we first reveal the underlying mechanism of spin degeneracy lifting in rhombohedral trilayer graphene. Then, we propose a design concept for all-rhombohedral graphene-based magnetic tunnel junctions (MTJs) by utilizing pristine, back-gated, and top-gated ABC-stacked trilayer graphene, which exhibit semimetallic (conducting), semiconducting (insulating), and half-metallic (ferromagnetic) behavior, respectively. This enables the realization of an "all-in-one" magnetic tunnel junction based entirely on trilayer graphene. This design enables voltage-controlled spintronics (lower power than conventional MTJs) with perfect interfacial matching and sub-nm thickness uniformity across 4-inch wafers. Using first-principles calculations and the non-equilibrium Green’s function, we comprehensively study electronic structures and transport properties of these all-graphene MTJs. Furthermore, we demonstrate that their characteristics can be tuned via a perpendicular electric field and electron doping. Our findings offer a new concept for the development of fully graphene-based spintronic devices utilizing the three distinct electronic phases of rhombohedral trilayer graphene.
\end{abstract}

\maketitle


Graphene is a zero-gap semiconductor, and much of the current research focuses on methods to induce a bandgap for applications in electronic devices \cite{novoselov2004electric}. Extensive works have been conducted on multiple-layer graphene. For example, many bilayer graphene works reveal that a tunable bandgap can be opened by applying an external electric field or through doping \cite{doi:10.1126/science.1130681,PhysRevLett.99.216802,PhysRevB.78.235408,zhang2009direct}. For few-layer graphene, such as trilayer, four-layer, and five-layer structures, the electronic properties depend strongly on the stacking order. In particular, non-identical stacking configurations result in an energy spectrum with bands exhibiting cubic dispersion and minimal curvature near the Fermi level. This flattening of bands, along with an enhanced density of states, indicates a significant role of many-body interactions in this regime. Therefore, the diverse stacking configurations in few-layer graphene give rise to a wide range of exotic properties, including strongly correlated electronic states \cite{zhang2011experimental,lui2011observation,han2024correlated,liu2024spontaneous}, the anomalous Hall effect \cite{chen2023gate}, quantized anomalous Hall effect, superconductivity, fluctuating magnetism, the Pomeranchuk effect, inter-valley coherent order, half-metallicity, and spin-orbit coupling \cite{patterson2025superconductivity,holleis2025fluctuating,choi2025superconductivity,arp2024intervalley,kim2023imaging,zhou2021superconductivity,zhou2021half}.

\begin{figure*}[ht!]
\centering
\includegraphics[width=0.9\textwidth]{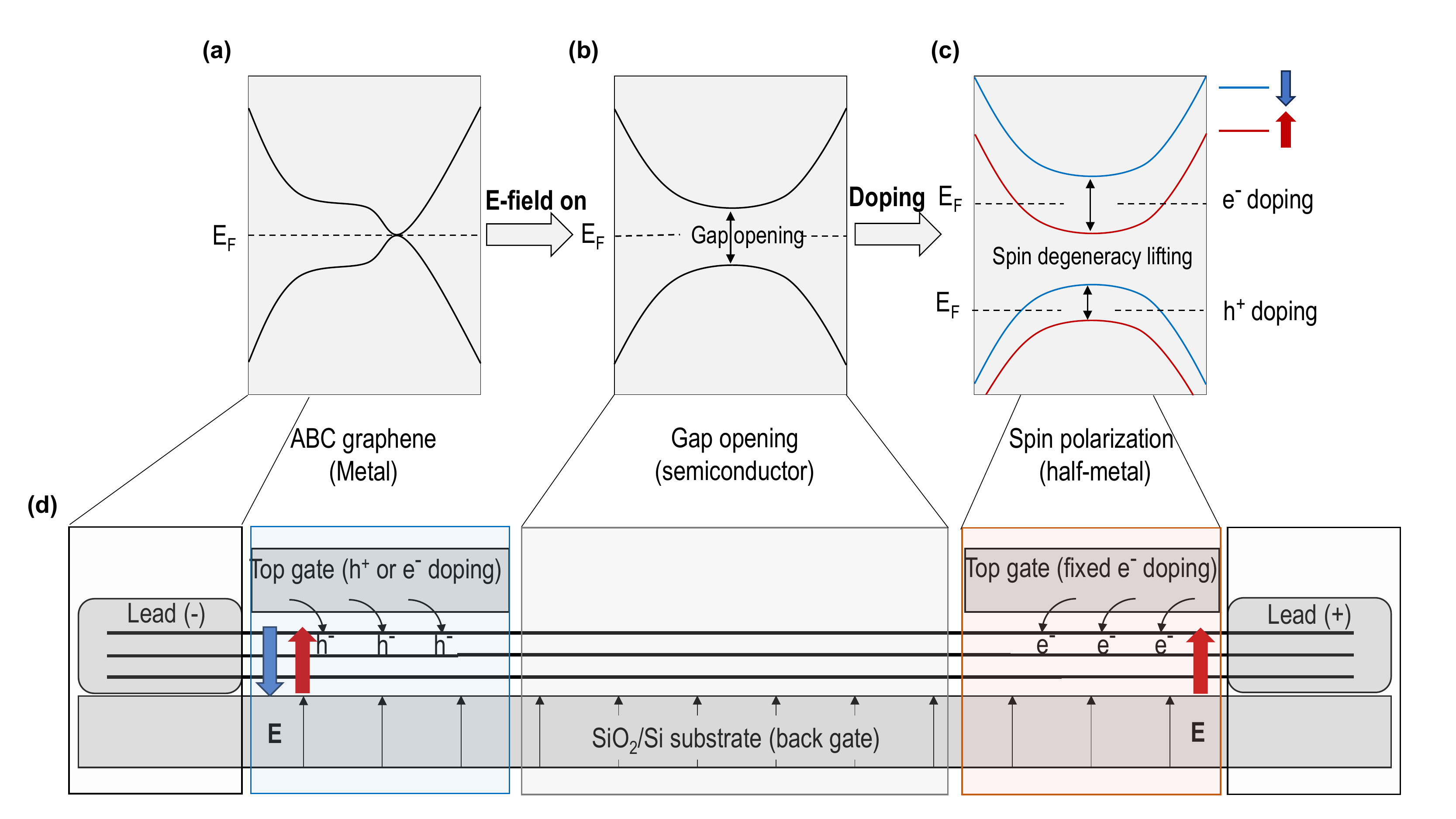}
\captionsetup{name=Schematic} 
\caption*{SCHEMATIC: Schematic illustration of the device design concept based on all ABC-stacked trilayer graphene. (a) Band structure of metallic, pristine rhombohedral trilayer graphene, which serves as the lead material in magnetic tunnel junctions. (b) Band structure of rhombohedral graphene under a back gate, which opens a semiconducting gap, enabling its use as the tunneling barrier in MTJs. (c) Band structure of rhombohedral graphene under both a back gate and a top gate. The top gate injects either electrons or holes, lifting the two-fold spin degeneracy and resulting in spin-up or spin-down polarization, respectively. This half-metallic behavior enables its use as a spin filter, allowing electrical switching of spin polarization at one end to realize either parallel (P) or anti-parallel (AP) magnetic configurations. (d) Structural configuration of the MTJ device entirely based on rhombohedral trilayer graphene.}
\label{fig:schematic}
\end{figure*}

Among the various stacking configurations of this two-dimensional (2D) material, rhombohedral (ABC) trilayer graphene holds a special position due to its broken inversion symmetry. This characteristic, where the graphene layers are stacked in a specific sequential order, significantly influences the electronic band structure, leading to the emergence of flat bands at the Fermi level \cite{henck2018flat, wang2018flat, aoki2007dependence}. These flat bands are known to enhance electron-electron interactions \cite{henck2018flat, koshino2010interlayer}. Recently, experimental advancements have revealed that ABC-stacked graphene can exhibit substantial spin splitting even in the absence of an external magnetic field, with phenomena such as fluctuating magnetism and the anomalous Hall effect being observed \cite{holleis2025fluctuating, choi2025superconductivity}. These effects are attributed to the intrinsic symmetry properties of the electronic structure and the Stoner mechanism at the Fermi level. Experimentally, Liu et al. \cite{liu2024spontaneous} investigated the role of Coulomb interactions in rhombohedral graphene, showing that these interactions can spontaneously break various symmetries. They demonstrated a transition from a layer-antiferromagnetic insulator with a 15 meV gap to a layer-polarized insulator by varying the electric displacement field. Subsequently, Huang et al. \cite{huang2023spin} provided a theoretical explanation for the metallic, spin-valley symmetry-broken states observed in hole-doped rhombohedral trilayer graphene under strong electric displacement fields. They emphasized the importance of momentum-space condensation in reconstructing the Fermi surface, driven by flavor symmetry breaking. Further modulation of carrier density and displacement fields enabled the exploration of isospin-polarized metals, including spin-valley-polarized and spin-polarized metallic states. These findings underscore the pivotal role of Coulomb interactions in inducing symmetry breaking in layered graphene systems. Such emergent physical phenomena not only simplify device architecture but also enhance spin manipulation efficiency, making ABC-stacked graphene a promising candidate for next-generation spintronic applications.

From a device perspective, the potential of graphene in giant magnetoresistance (GMR) applications has been emphasized due to its sensitivity to spin-polarized currents. These currents can be modulated by changes in magnetic alignment, resulting in variations in electrical resistance \cite{zhou2021half, zhou2021superconductivity, zhou2022isospin}. GMR effects are critical for developing magnetic sensors and memory storage devices. However, GMR has fundamental limitations, and tunneling magnetoresistance (TMR) technology is considered superior for memory applications owing to its higher signal-to-noise ratio, better scalability, and greater energy efficiency, which are all essential for next-generation non-volatile memory. Notably, all-graphene-based TMR devices have yet to be reported as it is hard to achieve all metallic, semiconducting, and ferromagnetic states within the same structure of graphene.  Rhombohedral-stacked graphene addresses this challenge by enabling tunable spin-split band structures essential for TMR implementation, while recent advances in chemical vapor deposition demonstrate wafer-scale integration pathways for such stacked 2D material systems \cite{liu2025homoepitaxial}.

In this work, we employ first-principles calculations to investigate spin polarization in ABC-stacked graphene under various electronic and magnetic conditions. Our study aims to understand how controlled electric fields and electron doping affect spin polarization and magnetoresistive responses, as illustrated in Schematic a–c. Based on these insights, we design all-ABC-stacked graphene magnetic tunnel junctions (as shown in Schematic d) and analyze their electronic transport properties using the non-equilibrium Green’s function method.


The first principle calculations are done using the Quantum-ATK package.\cite{ni2013transport, zhang2019taper}. The structure is relaxed till forces on each atom go below 0.005 eV/\AA. In our work, we use a rectangular unit cell (Fig.\ref{fig:fig1}a) which is more convenient for transport calculations. For self consistent calculation, we use spin polarised SG15 pseudo potential with a density mesh cutoff of 40 Hartree. In addition, we use an onsite Hubbard U parameter of 7.44 eV which is obtained from the constrained random phase approximation as implemented in code VASP \cite{Kresse1996}. The correction due to van-der Waals force is introduced via DFT-D2 method. For the band structure calculation we use a $30 \times 52$ k-mesh and a density mesh cut off of 40 Hartree.

To study the transport properties, we construct a device configuration by extending our unit cell along $z$ axis which we consider as the transport direction. We consider a scattering region of 22.14~\AA\ attached to two semi-infinite electrode. The semi-infinite electrodes are replaced by their self-energies which is calculated using recursion method as implemented in Quantum-ATK package. This is an iterative method which gives faster convergence. For the non-equlibrium calculation we use a $15 \times 1 \times 300$ k-mesh where values corresponds the sampling along the $x$, $y$ and $z$ direction. We use a tolerance of 0.0001 to ensure the convergence of the self consistent cycle.

\setcounter{figure}{0}
\begin{figure}[htbp]
\centering
\includegraphics[width=0.9\linewidth]{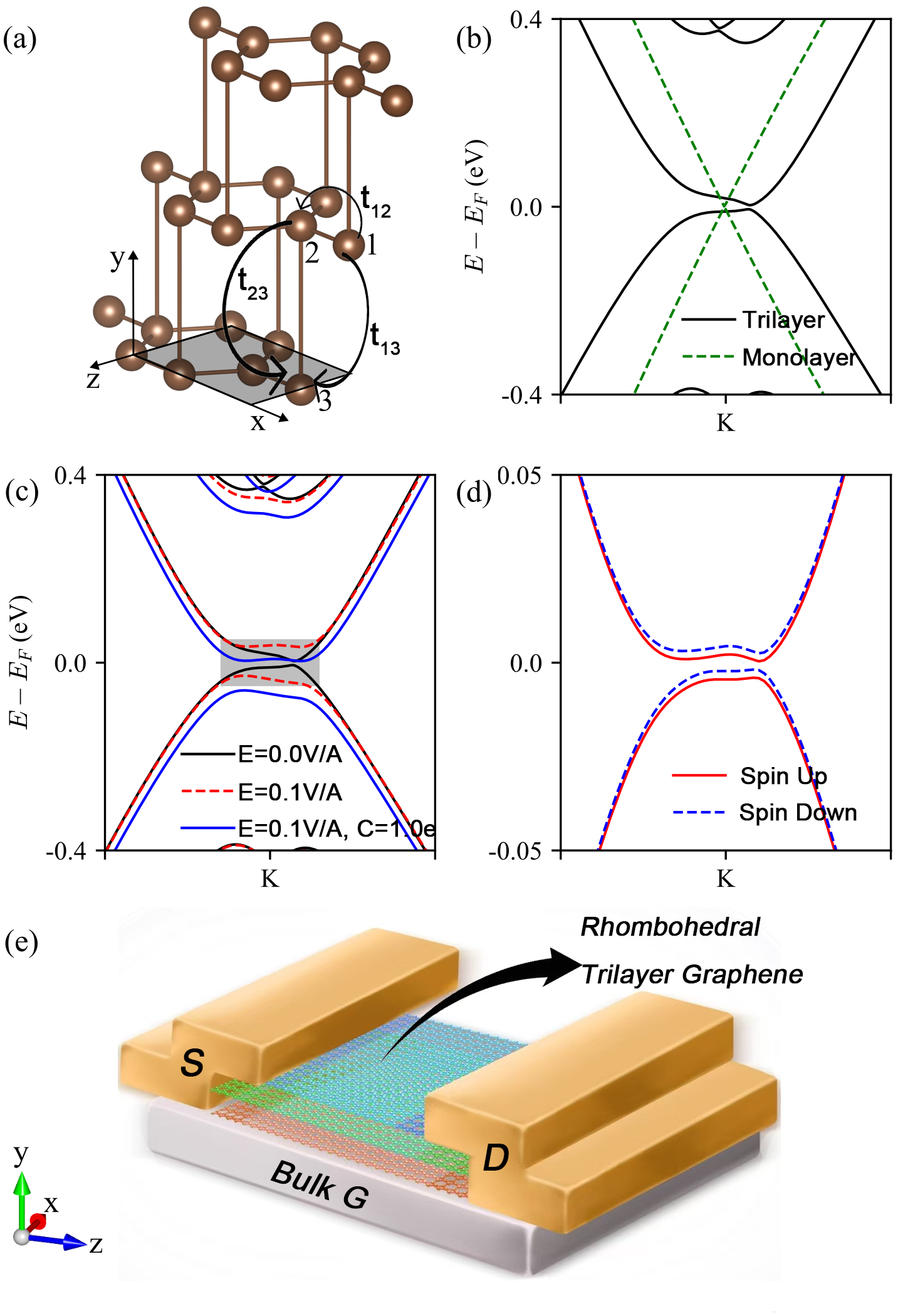}
\caption{(a) Crystal structure of ABC stacked graphene trilayer. Grey box shows the area of the unit cell. (b) Non spin polarised band structure of ABC trilayer (black solid line). Green dashed line show the band structure of a graphene monolayer for comparison. (c) Band structure with a gate voltage of 0.1 eV applied between two metallic gates separated by 20 \AA\,along $y$ direction (red dashed lines). The Blue lines show the band structure when an extra electron is added to the system. Black lines show the free trilayer band structure as (b). (d) Zoomed in spin polarised band structure showing the exchange splitting where red and blue dashed line show the up and down spin. Corresponding region is shown in (c) with a grey rectangle. (e) Two terminal device configuration used for transport calculation where green and red sites show the left and the right electrodes.}
\label{fig:fig1}
\end{figure}

The zero bias transmission coefficient of the device is obtained from
\begin{equation}
 T(E) = \mathrm{Tr}[\Gamma_L(E)G^{R}(E)\Gamma_R(E)G^{A}(E)]
\end{equation}
where $G^{R}(E)$  is the retarded Green’s functions and $G^{A}(E)$  is the advanced Green’s functions of the scattering region.   $\Gamma_L(E)= i(
\sum_{L(R)}^R-\sum_{L(R)}^A)$ is a coupling function between the structure and the left (right) electrode, and  $\sum_{L(R)}^{R(A)}$ is the retarded(advanced) self-energy matrix of the left and right semi-infinite electrodes. The total current due to an applied bias voltage can be obtained from the transmission coefficient using
\begin{equation}
I = \frac{e}{h} \int_{\mu_R}^{\mu _L} T(E) dE
\end{equation}
where  $\mu_R$ and  $\mu_L$ are the chemical potentials of the right and left electrodes respectively, and \( \mu_L - \mu_R = eV_B \), with \textit{V} being the bias voltage. In case of the spin polarised transport, one can further define the magneto-resistance (MR) as 
\begin{eqnarray}
MR = \frac{I_{\uparrow} - I_{\downarrow}}{I_{\uparrow} + I_{\downarrow}}.
\end{eqnarray}  
where $I_{\uparrow,\downarrow}$ correspond to current from the spin $\uparrow,\downarrow$ channel. For a small bias voltage, when the system can be assumed to be within the linear response regime and the total current can be approximated by $I = T(E_F)eV_B$. In such case the one can also define  $MR = (T_{\uparrow} - T_{\downarrow})/(T_{\uparrow} + T_{\downarrow})$.


In an ABC-stacked trilayer graphene (Fig.\ref{fig:fig1}a), each layer is shifted along the bond by the bond length itself (see Fig.\ref{fig:fig1}a). Our first principles calculation shows that the bond length between two C atoms in the same plane is 1.42 \AA \,and the interlayer distance is 3.6 \AA. Contrary to AB stacked bilayer graphene, in an ABC stacked structure, the band gap tends to become zero \cite{Zhang2010} (Fig.\ref{fig:fig1}b). However the point where the valence and conduction band approaches each other is shifted from $K/K'$ point which is observed when compared to the band structure of the monolayer graphene. Presence of an electric field perpendicular to the plane of the layer breaks the sublattice degeneracy and increase the bandgap (Fig.\ref{fig:fig1}c). Note that, with the gap opening the bands have become significantly flat \cite{Wang2013, henck2018flat} which makes them suitable candidate for the Coulomb interaction and consequently for magnetism. It is because the presence of the onsite Hubbard interaction can result in a spin splitting (Fig.\ref{fig:fig1}d), which is more prominent in the surface C atoms that dominate states near the Fermi level. This can be observed by the exchange splitting of the bands (Fig.\ref{fig:fig1}d). For U=7.44 eV we observe an exchange splitting of $\sim$2.2 meV near K(K') point.

\begin{figure}[ht!]
\centering
\includegraphics[width=1\linewidth]{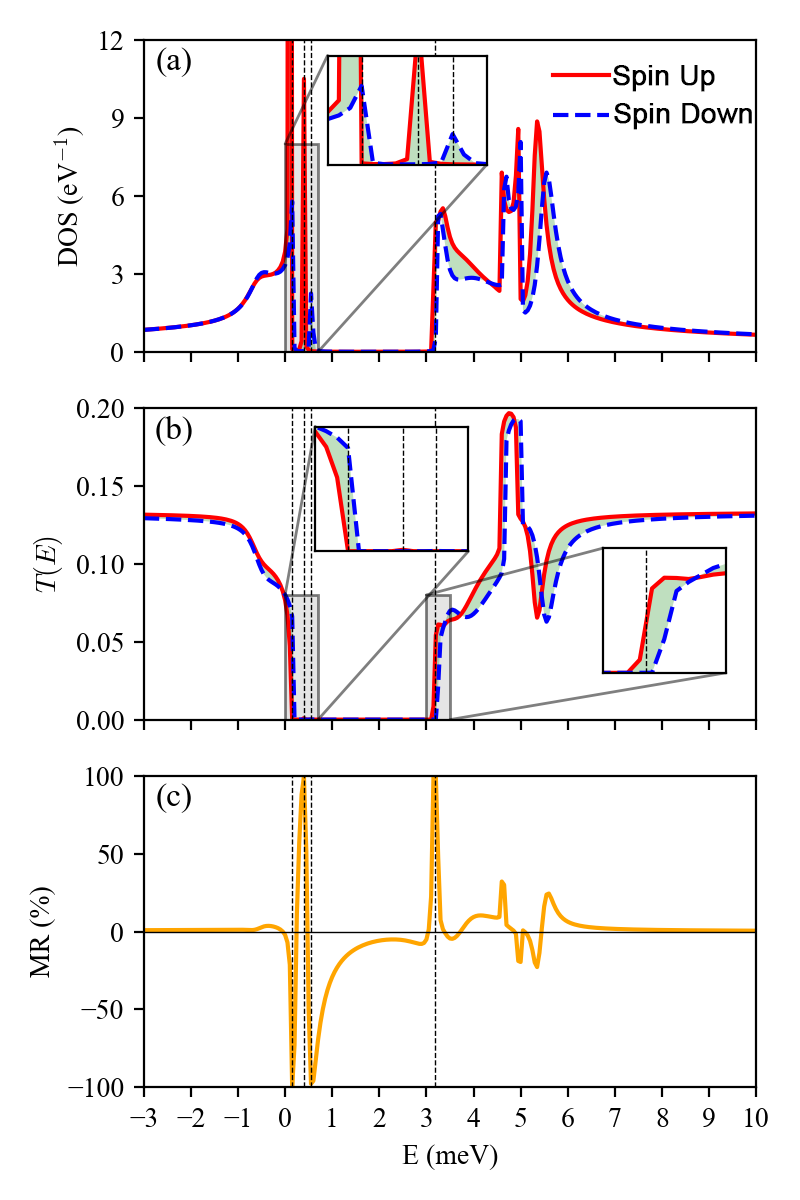}
\caption{(a) Spin-polarized density of states for trilayer graphene. (b) Spin-polarized transmission spectrum. Insets show the zoomed-in regions of the band edges. (c) Magnetoresistance (in $\%$) for the trilayer graphene. Vertical dashed lines show the positions where the MR reaches 100\%.}
\label{fig:fig2}
\end{figure}

The spin splitting at the valence band maxima and conduction band minima (Fig.\ref{fig:fig1}d) provides an unique opportunity to obtain fully spin polarised states which is not possible with normal metallic conductors. Note that, compared to the monolayer graphene which demonstrates massless Dirac cone,  the flatness of the bands from the trilayer graphene reduces the group velocity of the electrons. As a result, near the band edges one can observe peaks in density of states (DOS) which is coming from the localised states (Fig.\ref{fig:fig2}a). Note that large value of DOS at Fermi level also satisfies the so-called Stoner criterion \cite{stoner1938collective} from the mean-field solution of the Hubbard model. For an Hubbard parameter $U$=7.44 eV, one can readily see that the DOS satisfies the condition $UN(E_F) > 1$, where $N(E_F)$=($N^\uparrow(E_F)$ + $N^\downarrow(E_F)$) is the total density of states at the Fermi level, ensuring the ferromagnetic nature of the trilayer graphene.

The physical origin of the ferromagnetism in trilayer graphene can be further studied by constructing a generalized Hubbard model
\begin{equation}
H = \sum_{\langle ij \rangle \sigma} t_{ij}^{\sigma} c^{\dagger}_{i\sigma} c_{j\sigma} + \sum_{\langle ijkl \rangle \sigma \sigma'} U_{ijkl}^{\sigma \sigma'} c^{\dagger}_{i\sigma} c^{\dagger}_{k\sigma'} c_{j\sigma} c_{k\sigma'} 
\end{equation}
The Hubbard $U$ and exchange $J$ parameters are calculated from the $U_{ijkl}^{\sigma \sigma'}$ parameters \cite{Vaugier2012} using constrained random phase approximation with ab-initio code VASP \cite{Kresse1996} using three conduction bands. Here, we use the carbon atom at the middle graphene layer as an example. Index 1,2 correspond nearest carbon atom in same layer while 2,3 represent the nearest carbon atoms from different layers (Fig.\ref{fig:fig1}a). Our simulation leads to U$_{22}$ = 8.22 eV, U$_{12}$ = 2.63 eV, U$_{23}$ = 1.63 eV, J$_{12}$ = 2.83 meV, J$_{23}$ = 0.13 meV, t$_{12}$ = 2.91, t$_{23}$ = 0.046 and t$_{13}$ = 0.0089, where the U$_{22}$ is the onsite and U$_{21}$,U$_{23}$ are the off-site Coulomb energy. J$_{12}$, J$_{23}$ are the direct exchange. t$_{12,23,13}$ are the hopping integrals (Fig.\ref{fig:fig1}a). The large on-site Coulomb interaction indicates the strong localization of this $p_z$ orbital. The relatively large off-site Coulomb interaction shows the moderate screening effects from the in-plane and out-of-plane carbon atoms. The much large direction exchange between the carbon 2 and 1 than that between 2 and 3 suggests the larger orbital overlap of between the in-plane carbon atoms than the out-of-planes ones. This is also reflected by the order of the hopping integrals magnitudes by $t_{12} > t_{23} > t_{13}$.

After achieving the spin-polarized state, we construct a
two terminal device as shown in Fig.\ref{fig:fig1}e and calculate the
transmission coefficient and magnetoresistance of the devices. One can construct a magnetic tunnel junction by combining similar configuration with different magnetic alignment (Fig.Schematic). The transmission coefficient in Fig.\ref{fig:fig2}b shows a non-monotonic behavior near the band edges which eventually saturates to a constant value. Most interestingly, near the band edges there are states which provide fully spin-polarized conduction channel (Fig.\ref{fig:fig2}b). This can be explained by the band structure in Fig.\ref{fig:fig1}. This behavior becomes more important when we calculate the MR to estimate the applicability of the trilayer graphene as MTJs (Fig.\ref{fig:fig2}c). Due to the presence of fully spin-polarized conduction channels near the band edges, the MR reaches 100$\%$ with an opposite sign for valence and conduction band edges (denoted by extreme left and right vertical dashed lines in Fig.\ref{fig:fig2}c). Note that there are other peaks denoting  almost 100$\%$ MR close to valence band edge. Careful observation will show that at these energies, the transmission coefficient corresponding to both spin channels are almost zero (Fig.\ref{fig:fig2}b), however a specific spin has a strong peak in DOS (Fig.\ref{fig:fig2}a). This is an indication of the tunneling state which carries very small current. In spite of the smaller magnitude of TMR, the states which are away from the Fermi level carry more current and therefore can be easier to access and manipulate for the transport applications since they possess moderately large net transmission. These states can brought close to the Fermi level with the help of gate voltage and external doping which provides an unique opportunity to utilize both the tunneling and transmissive states depending on the requirement.

In conclusion, our study reveals that the electronic and spintronic properties of ABC-stacked trilayer graphene can be exquisitely tuned through the strategic manipulation of electron doping and external electric fields. We demonstrate that such control facilitates the distinct separation of spin-up and spin-down states, primarily through the electron-electron interaction, which is manifested strongly near K/K' point of the Brillouin zone. As a result, one can obtain fully spin-polarized current resulting a 100$\%$ magnetoresistance. In addition, we also observe 100$\%$ magnetoresistance due to tunneling through the localized states which we excluded by analyzing the density of states.  These findings underscore the substantial potential of graphene for technologies that demand precise control over electronic properties and spin states. The ability to shift the Fermi level via electron doping, thereby altering the band curvature and transitioning the material between semiconducting and metallic states, further elucidates the dual utility of electric fields and doping in engineering the electronic landscape of graphene. This research not only deepens our understanding of two-dimensional material physics but also enhances the prospects for graphene-based devices in applications ranging from quantum computing to advanced spintronics, where the manipulation of spin-polarized currents is critical.


\begin{acknowledgments}
S.G. is co-funded by the European Union (Physics for Future – Grant Agreement No. 101081515). L. S. acknowledges the Ministry of Education, Singapore (Grant No. A-8001194-00-00 and Grant No. A-8001872-00-00). J. Yang acknowledges the National Natural Science Foundation of China (No. 12404286), the Key Scientific Research Project of Colleges and Universities in He'nan Province (No. 24A140022), and the Natural Science Foundation of He'nan Province (No. 242300421671). Y.M. Zhao gratefully acknowledges the support of China Scholarship Council (No. CSC202104910079).
\end{acknowledgments}

\bibliographystyle{unsrt}
\bibliography{main}

\begin{thebibliography}{10}

\bibitem{novoselov2004electric}
Kostya~S Novoselov, Andre~K Geim, Sergei~V Morozov, De-eng Jiang, Yanshui Zhang, Sergey~V Dubonos, Irina~V Grigorieva, and Alexandr~A Firsov.
\newblock Electric field effect in atomically thin carbon films.
\newblock {\em science}, 306(5696):666--669, 2004.

\bibitem{doi:10.1126/science.1130681}
Taisuke Ohta, Aaron Bostwick, Thomas Seyller, Karsten Horn, and Eli Rotenberg.
\newblock Controlling the electronic structure of bilayer graphene.
\newblock {\em Science}, 313(5789):951--954, 2006.

\bibitem{PhysRevLett.99.216802}
Eduardo~V. Castro, K.~S. Novoselov, S.~V. Morozov, N.~M.~R. Peres, J.~M. B.~Lopes dos Santos, Johan Nilsson, F.~Guinea, A.~K. Geim, and A.~H.~Castro Neto.
\newblock Biased bilayer graphene: Semiconductor with a gap tunable by the electric field effect.
\newblock {\em Phys. Rev. Lett.}, 99:216802, Nov 2007.

\bibitem{PhysRevB.78.235408}
L.~M. Zhang, Z.~Q. Li, D.~N. Basov, M.~M. Fogler, Z.~Hao, and M.~C. Martin.
\newblock Determination of the electronic structure of bilayer graphene from infrared spectroscopy.
\newblock {\em Phys. Rev. B}, 78:235408, Dec 2008.

\bibitem{zhang2009direct}
Yuanbo Zhang, Tsung-Ta Tang, Caglar Girit, Zhao Hao, Michael~C Martin, Alex Zettl, Michael~F Crommie, Y~Ron Shen, and Feng Wang.
\newblock Direct observation of a widely tunable bandgap in bilayer graphene.
\newblock {\em Nature}, 459(7248):820--823, 2009.

\bibitem{zhang2011experimental}
Liyuan Zhang, Yan Zhang, Jorge Camacho, Maxim Khodas, and Igor Zaliznyak.
\newblock The experimental observation of quantum hall effect of l= 3 chiral quasiparticles in trilayer graphene.
\newblock {\em Nature Physics}, 7(12):953--957, 2011.

\bibitem{lui2011observation}
Chun~Hung Lui, Zhiqiang Li, Kin~Fai Mak, Emmanuele Cappelluti, and Tony~F Heinz.
\newblock Observation of an electrically tunable band gap in trilayer graphene.
\newblock {\em Nature Physics}, 7(12):944--947, 2011.

\bibitem{han2024correlated}
Tonghang Han, Zhengguang Lu, Giovanni Scuri, Jiho Sung, Jue Wang, Tianyi Han, Kenji Watanabe, Takashi Taniguchi, Hongkun Park, and Long Ju.
\newblock Correlated insulator and chern insulators in pentalayer rhombohedral-stacked graphene.
\newblock {\em Nature Nanotechnology}, 19(2):181--187, 2024.

\bibitem{liu2024spontaneous}
Kai Liu, Jian Zheng, Yating Sha, Bosai Lyu, Fengping Li, Youngju Park, Yulu Ren, Kenji Watanabe, Takashi Taniguchi, Jinfeng Jia, et~al.
\newblock Spontaneous broken-symmetry insulator and metals in tetralayer rhombohedral graphene.
\newblock {\em Nature nanotechnology}, 19(2):188--195, 2024.

\bibitem{chen2023gate}
Hao Chen, Arpit Arora, Justin~CW Song, and Kian~Ping Loh.
\newblock Gate-tunable anomalous hall effect in bernal tetralayer graphene.
\newblock {\em Nature Communications}, 14(1):7925, 2023.

\bibitem{patterson2025superconductivity}
Caitlin~L Patterson, Owen~I Sheekey, Trevor~B Arp, Ludwig~FW Holleis, Jin~Ming Koh, Youngjoon Choi, Tian Xie, Siyuan Xu, Yi~Guo, Hari Stoyanov, et~al.
\newblock Superconductivity and spin canting in spin--orbit-coupled trilayer graphene.
\newblock {\em Nature}, pages 1--7, 2025.

\bibitem{holleis2025fluctuating}
Ludwig Holleis, Tian Xie, Siyuan Xu, Haoxin Zhou, Caitlin~L Patterson, Archisman Panigrahi, Takashi Taniguchi, Kenji Watanabe, Leonid~S Levitov, Chenhao Jin, et~al.
\newblock Fluctuating magnetism and pomeranchuk effect in multilayer graphene.
\newblock {\em Nature}, 640(8058):355--360, 2025.

\bibitem{choi2025superconductivity}
Youngjoon Choi, Ysun Choi, Marco Valentini, Caitlin~L Patterson, Ludwig~FW Holleis, Owen~I Sheekey, Hari Stoyanov, Xiang Cheng, Takashi Taniguchi, Kenji Watanabe, et~al.
\newblock Superconductivity and quantized anomalous hall effect in rhombohedral graphene.
\newblock {\em Nature}, pages 1--6, 2025.

\bibitem{arp2024intervalley}
Trevor Arp, Owen Sheekey, Haoxin Zhou, CL~Tschirhart, Caitlin~L Patterson, HM~Yoo, Ludwig Holleis, Evgeny Redekop, Grigory Babikyan, Tian Xie, et~al.
\newblock Intervalley coherence and intrinsic spin--orbit coupling in rhombohedral trilayer graphene.
\newblock {\em Nature Physics}, 20(9):1413--1420, 2024.

\bibitem{kim2023imaging}
Hyunjin Kim, Youngjoon Choi, {\'E}tienne Lantagne-Hurtubise, Cyprian Lewandowski, Alex Thomson, Lingyuan Kong, Haoxin Zhou, Eli Baum, Yiran Zhang, Ludwig Holleis, et~al.
\newblock Imaging inter-valley coherent order in magic-angle twisted trilayer graphene.
\newblock {\em Nature}, 623(7989):942--948, 2023.

\bibitem{zhou2021superconductivity}
Haoxin Zhou, Tian Xie, Takashi Taniguchi, Kenji Watanabe, and Andrea~F Young.
\newblock Superconductivity in rhombohedral trilayer graphene.
\newblock {\em Nature}, 598(7881):434--438, 2021.

\bibitem{zhou2021half}
Haoxin Zhou, Tian Xie, Areg Ghazaryan, Tobias Holder, James~R Ehrets, Eric~M Spanton, Takashi Taniguchi, Kenji Watanabe, Erez Berg, Maksym Serbyn, et~al.
\newblock Half-and quarter-metals in rhombohedral trilayer graphene.
\newblock {\em Nature}, 598(7881):429--433, 2021.

\bibitem{henck2018flat}
Hugo Henck, Jose Avila, Zeineb Ben~Aziza, Debora Pierucci, Jacopo Baima, Bet{\"u}l Pamuk, Julien Chaste, Daniel Utt, Miroslav Bartos, Karol Nogajewski, et~al.
\newblock Flat electronic bands in long sequences of rhombohedral-stacked graphene.
\newblock {\em Physical Review B}, 97(24):245421, 2018.

\bibitem{wang2018flat}
Weimin Wang, Yuchen Shi, Alexei~A Zakharov, Mikael Syvajarvi, Rositsa Yakimova, Roger~IG Uhrberg, and Jianwu Sun.
\newblock Flat-band electronic structure and interlayer spacing influence in rhombohedral four-layer graphene.
\newblock {\em Nano Letters}, 18(9):5862--5866, 2018.

\bibitem{aoki2007dependence}
Masato Aoki and Hiroshi Amawashi.
\newblock Dependence of band structures on stacking and field in layered graphene.
\newblock {\em Solid State Communications}, 142(3):123--127, 2007.

\bibitem{koshino2010interlayer}
Mikito Koshino.
\newblock Interlayer screening effect in graphene multilayers with a b a and a b c stacking.
\newblock {\em Physical Review B}, 81(12):125304, 2010.

\bibitem{huang2023spin}
Chunli Huang, Tobias~MR Wolf, Wei Qin, Nemin Wei, Igor~V Blinov, and Allan~H MacDonald.
\newblock Spin and orbital metallic magnetism in rhombohedral trilayer graphene.
\newblock {\em Physical Review B}, 107(12):L121405, 2023.

\bibitem{zhou2022isospin}
Haoxin Zhou, Ludwig Holleis, Yu~Saito, Liam Cohen, William Huynh, Caitlin~L Patterson, Fangyuan Yang, Takashi Taniguchi, Kenji Watanabe, and Andrea~F Young.
\newblock Isospin magnetism and spin-polarized superconductivity in bernal bilayer graphene.
\newblock {\em Science}, 375(6582):774--778, 2022.

\bibitem{liu2025homoepitaxial}
Lei Liu, Taotao Li, Xiaoshu Gong, Hengdi Wen, Liqi Zhou, Mingwei Feng, Haotian Zhang, Ningmu Zou, Shengqiang Wu, Yuhao Li, et~al.
\newblock Homoepitaxial growth of large-area rhombohedral-stacked mos2.
\newblock {\em Nature Materials}, pages 1--8, 2025.

\bibitem{ni2013transport}
Yun Ni, Kai-Lun Yao, Hua-Hua Fu, Guo-Ying Gao, Si-Cong Zhu, Bo~Luo, Shu-Ling Wang, and Rui-Xue Li.
\newblock The transport properties and new device design: the case of 6, 6, 12-graphyne nanoribbons.
\newblock {\em Nanoscale}, 5(10):4468--4475, 2013.

\bibitem{zhang2019taper}
Lishu Zhang, Jiaren Yuan, Lei Shen, Cameron Fletcher, Xingfan Zhang, Tao Li, Xinyue Dai, Yanyan Jiang, and Hui Li.
\newblock Taper-shaped carbon based spin filter.
\newblock {\em Applied Surface Science}, 495:143501, 2019.

\bibitem{Kresse1996}
G.~Kresse and J.~Furthm\"uller.
\newblock Efficient iterative schemes for ab initio total-energy calculations using a plane-wave basis set.
\newblock {\em Phys. Rev. B}, 54:11169--11186, Oct 1996.

\bibitem{Zhang2010}
Fan Zhang, Bhagawan Sahu, Hongki Min, and A.~H. MacDonald.
\newblock Band structure of $abc$-stacked graphene trilayers.
\newblock {\em Phys. Rev. B}, 82:035409, Jul 2010.

\bibitem{Wang2013}
Hao Wang, Jin-Hua Gao, and Fu-Chun Zhang.
\newblock Flat band electrons and interactions in rhombohedral trilayer graphene.
\newblock {\em Phys. Rev. B}, 87:155116, Apr 2013.

\bibitem{stoner1938collective}
Edmund~Clifton Stoner.
\newblock Collective electron ferromagnetism.
\newblock {\em Proceedings of the Royal Society of London. Series A. Mathematical and Physical Sciences}, 165(922):372--414, 1938.

\bibitem{Vaugier2012}
Lo\"{\i}g Vaugier, Hong Jiang, and Silke Biermann.
\newblock Hubbard $u$ and hund exchange $j$ in transition metal oxides: Screening versus localization trends from constrained random phase approximation.
\newblock {\em Phys. Rev. B}, 86:165105, Oct 2012.

\end{thebibliography}

\end{document}